\documentclass[useAMS,usenatbib]{mn2e}
\usepackage{graphicx, rotating}
\usepackage{aas_macros}  
\usepackage{graphicx}
\usepackage{amssymb}
\usepackage{rotating}
\usepackage{multirow}
\usepackage[breaklinks,colorlinks=true,citecolor=black,linkcolor=black,urlcolor=blue,bookmarks=true]{hyperref}
\usepackage{breakurl}  

\newcommand{\csq}{$\chi^2$}
\newcommand{\kms}{km\,s$^{-1}$}

\newcommand{\msol}{\textrm{M}$_{\odot}$}
\newcommand{\lsol}{\textrm{L}$_{\odot}$}
\newcommand{\hi}{\textrm{H}{\sc i}}
\newcommand{\hii}{\textrm{H}{\sc ii}}

\title[KDG\,61 \& 64 star formation history]{Star formation history of 
KDG\,61 and KDG\,64 from spectroscopy and colour-magnitude diagrams\thanks{
Based on observations made with the NASA/ESA Hubble Space Telescope,
obtained from the data archive at the Space Telescope Science Institute.
STScI is operated by the Association of Universities for Research
in Astronomy, Inc. under NASA contract NAS 5-26555.}
\thanks{Based on observations made with the 6m BTA telescope of
the Special Astrophysical Observatory, Russian Academy of Sciences.}
}

\author[Makarova et al.]{
Lidia Makarova$^{1,2}$\thanks{E-mail: lidia@sao.ru}, 
Mina Koleva$^{3,2}$,  
Dmitry Makarov$^{1,2}$, 
Philippe Prugniel$^{2}$\\
$^{1}$Special Astrophysical Observatory, Nizhniy Arkhyz, 
Karachai-Cherkessia 369167, Russia\\
Isaac Newton Institute of Chile, SAO Branch, Russia\\
$^{2}$Universit\'e Lyon~1,
  Villeurbanne, F-69622, France; CRAL, Observatoire de Lyon, St. Genis
  Laval, F-69561, France ; \\ CNRS, UMR 5574 \\ 
$^{3}$Instituto de Astrof\'{\i}sica de Canarias, La Laguna, E-38200 Tenerife, Spain\\
Departamento de Astrof\'{\i}sica, Universidad de La Laguna, E-38205 La
Laguna, Tenerife, Spain\\ 
}

\begin{document}

\date{Accepted 2010 March 25. Received 2010 March 17; in original form 2010 January 22}

\pagerange{\pageref{firstpage}--\pageref{lastpage}} \pubyear{XXX}

\maketitle

\label{firstpage}

\begin{abstract}
A study of two dE/dSph members of the nearby M\,81
group of galaxies, KDG\,61 and UGC\,5442 = KDG\,64, has been made.
Direct Hubble Space Telescope (HST) Advanced Camera for Surveys (ACS) images
and integrated-light spectra of 6 m telescope of
Special Astrophysical Observatory of Russian Academy of Sciences
have been used for quantitative star formation history analysis.
The spectroscopic and colour-magnitude diagrams analysis gives consistent results.
These galaxies appear to be dominated by
an old population (12--14 Gyr) of low metallicity (${\rm [Fe/H]}\sim-1.5$).
Stars of ages about 1 to 4 Gyr have been detected in both galaxies.
The later population shows marginal metal enrichment.
We do not detect any
significant radial gradients in age or metallicity in these galaxies.
Our radial velocity measurement suggests that the \hii{} knot on
the line-of-sight of  KDG\,61 is not gravitationally attached to the
galaxy.

\end{abstract}


\section{Introduction}
Most of the known galaxies are dwarfs. They span mass ranges from a few
$10^9$ \msol, (or ${\rm M}_V \simeq -18$ mag; $10^9$ \lsol) like
NGC~205, to  $10^7$ \msol{} (${\rm M}_V \simeq -12$ mag; $10^6$
\lsol),  like Fornax and Sculptor galaxies, satellites of the
Milky Way, and probably  down to $10^5$ \msol{}  (${\rm M}_V \simeq
-3$ mag; 1000 \lsol) for the fainter Milky Way satellites found from
the SDSS \citep{koposov09}.  Dwarfs vary considerably in their gas
content and morphology, between dwarf irregular galaxies (dIrr) and
dwarf elliptical or spheroidal galaxies (dE/dSph).  Environmental
effects likely drive the evolution of star forming dwarf galaxies to
the quiescent ones after gas removal and/or  exhaustion.  Among the
quiescent objects, the fainter ones are traditionally called dSph
while the more massive are named dE (diffuse or dwarf ellipticals;
e.g. \citealp{grebel99}).  The bound is often set at ${\rm M}_V
\simeq -14$ mag.  However, the physical distinction between these two
classes is unclear. Dwarf spheroidals are generally companions of
massive galaxies (20 of them were identified around the Milky Way and 15
around Andromeda; \citealp{irwin08}) while dEs are found in clusters
(1141 of them are listed in the Virgo Cluster Catalogue;
\citealp{bst85}),  but NGC~205, prototype of dEs, is also a companion
of M\,31, and faint dE/dSph are now found in nearby clusters
\citep{trentham02,mieske07,adami07,derijcke09}.  The nearby dSphs are
often found to be dominated by dark matter \citep{mateo98}, while the
more massive, dEs are apparently similar to the massive elliptical
galaxies: the stellar content of NGC~147, NGC~185 and NGC~205 account
for half of their dynamical masses \citep{derijcke06}.  Other
properties, like the S\'ersic index, n, characterizing the shape of
the photometric profile, seems to form a continuum over all the mass
range. \cite{derijcke09} proposes that the S\'ersic index is
independent of the luminosity for ${\rm M}_V \gtrsim -14$ mag, $n
\approx 0.7$,  and increases with the luminosity above this limit. The
change in the photometric scaling relation may reflect the different
nature of dEs and dSphs, but both the measurement uncertainties and
the cosmic dispersion are still large enough to dispute such a
dichotomy.

Exploring the mass sequence of dEs/dSphs is a significant step
toward the understanding of the different processes of their formation 
and evolution. Is there a sharp transition of the dark matter content
at some intermediate mass? Were all these galaxies formed at the same
epoch and are there any systematics in the star formation?

To address these questions, we have studied two of the nearest
dE/dSph galaxies with intermediate masses, located in the M\,81 group 
at about 3.6\,Mpc. They are two of the brightest dE/dSph in this group: 
UGC\,5442 = KDG\,64
and KDG\,61. The KDG designations (Karachentseva dwarf galaxies) are 
from the galaxy catalogue by \citet{karachentseva68}.
 
Both objects are situated in the central region of the group.  KDG\,61
is located 29.5\,arcmin South of M\,81, i.e.\ 31\,kpc in  projected
distance.  KDG\,64 is projected 58.2\,arcmin South of NGC\,3077 (61.5\,kpc)  
and 97.5\,arcmin South-East of M\,81 (103\,kpc).  KDG\,61 is the
most luminous dSph in the M\,81 group  and it is one of the closest
companions to the M\,81 galaxy.  \citet{johnson97} has found a \hii{}
knot   34\,arcsec NE from the centre (0.6 kpc in projection,  position
J2000 09:57:07.48+68:35:53.9).  This sign of recent star formation and
the detection of a \hi{} cloud led \citet{k00} and \citet{boyce01} to
classify the galaxy as transitional type, between dIrr and dSph.  
A globular cluster has been discovered in KDG\,61 \citep{sharina05} at
the projected distance of 0.05\,kpc (3\,arcsec) from the galaxy's centre.
This cluster has an effective radius of 4.7\,pc (0.3\,arcsec), absolute
luminosity ${\rm M}_V\,=\,-7.55$ and a de-reddened  colour of $V-I\,=\,0.92$.
This colour is similar to the rest of the galaxy (see
Table~\ref{table:general}).  In spite of the fact that KDG\,64 is fainter than
KDG\,61, it has half  magnitude higher surface brightness.  The
nucleus-like object \citep{bp94,bremnes98}  is a distant
galaxy \citep{k00,sp02}.  No globular cluster was detected by
\citet{sharina05}.  The internal kinematics of KDG\,64 was studied
using long-slit  spectroscopy by \citet{sp02}. 

The absolute luminosity, total colour and other general parameters of
the both galaxies are shown  in the Table~\ref{table:general}.
Coordinates are taken from 
HyperLEDA\footnote{\url{http://leda.univ-lyon1.fr/}} \citep{patu03}.  
Total $B$ and $R$
magnitudes are taken from \citet{bremnes98}, central surface
brightness in $V$ and $(V-I)$ colour are from \citet{sharina08}.
The axial ratio was taken from the `A Catalog of Neighboring Galaxies'
\citep{CNG} and the central velocity dispersion from \citet{ps02}.
The rest of values are from this work. All magnitudes and colours are
corrected for Galactic extinction using the  \citet{schlegel98} maps.

\begin{table*}
\centering
\caption{General parameters of  KDG\,61 and KDG\,64}.
\label{table:general}
\begin{tabular}{lll}
\hline
                                        & KDG\,61                       & KDG\,64               \\
Position (J2000)                        & 095703.1$+$683531             & 100701.7$+$674938     \\
$E(B-V)$, mag                           & \phantom{$-$00}0.072          & \phantom{$-0$}0.054   \\
$B^0_T$, mag                            & \phantom{$-$0}14.93           & \phantom{$-$}15.29    \\
Axial ratio $b/a$                       & \phantom{$-$00}0.58           & \phantom{$-0$}0.47    \\
$(B-R)^0_T$, mag                        & \phantom{$-$00}1.44           & \phantom{$-0$}1.35    \\
$(V-I)^0_e$, mag                        & \phantom{$-$00}0.94           & \phantom{$-0$}0.99    \\
Heliocentric velocity, \kms{}           & \phantom{$-$}$221\pm3$        & $-15 \pm 13$      \\
Distance modulus $\mu(0)$, mag          & \phantom{$-$0}$27.77\pm0.04$  & \phantom{$-$}$27.84\pm0.04$       \\
Distance, Mpc                           & \phantom{$-$00}$3.58\pm0.07$  & \phantom{$-0$}$3.70\pm0.07$       \\
Spatial separation to M~81, kpc         & \phantom{$-$0}40              & 230                  \\
$B$ absolute magnitude, mag               & \phantom{0}$-12.84$           & $-12.55$          \\
Central surface brightness in $V$, mag arcsec$^{-2}$  &  \phantom{$-$0}23.76     & \phantom{$-$}22.89                \\
Central velocity dispersion, \kms{}     &                               & \phantom{$-$}$25.0 \pm  16.0$ \\
Fraction of old stars (12--14 Gyr)      & \phantom{$-$0}82--86 \%     & \phantom{$-$}72--89 \%                \\
Metallicity of old stars, [Fe/H], dex   & \phantom{00}$-1.5\pm0.2$      & \phantom{0}$-1.6\pm0.4$  \\
\hline
\end{tabular}
\end{table*} 

In Sect.\,2 we analyse colour-magnitude diagrams (CMDs) of KDG\,61 and KDG\,64
derived from HST/ACS images. We measure accurately the distance
and find the star formation and metal enrichment history.
In Sect.\,3 we present the long-slit spectra that we obtained and
their analysis using full spectrum fitting. Section\,4 discusses the
nature of these galaxies and Section\,5 gives our conclusions.

\section{Direct images and star formation history}

\subsection{Observational material and photometry}
\begin{figure*}
\includegraphics[width=8cm]{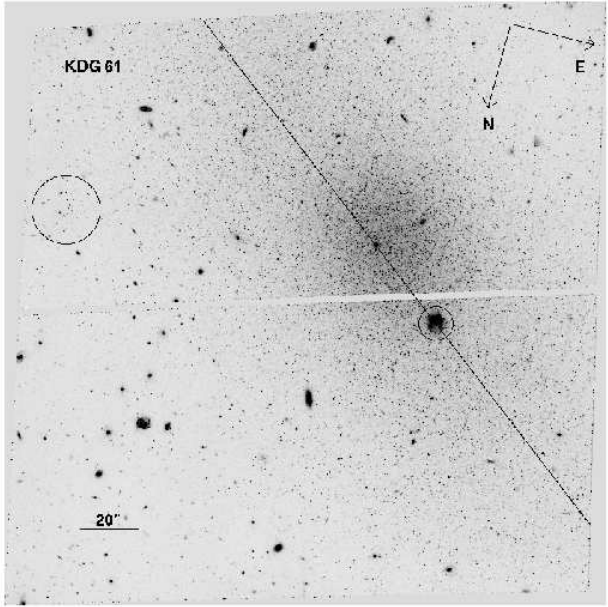}
\hspace{1.5cm}
\includegraphics[width=8cm]{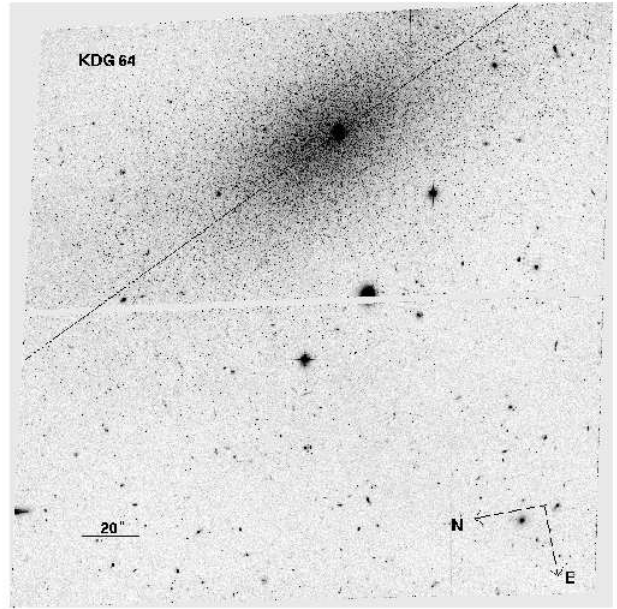}
\caption{HST/ACS images of  KDG\,61 and KDG\,64 in F606W filter. The position 
of the spectrograph slit is indicated with line. The slit width is 1 arcsec.
The \hii{} region and the loose concentration of blue stars are indicated
with circles (see Sect.~\ref{sect:blue}).} 
\label{fig:ima_spa}
\end{figure*}

Direct images of KDG\,61 and KDG\,64 were taken from the HST data
archive  (proposal 9884, PI Taft Armandroff).  Deep HST/ACS
observations of the  two galaxies were made using $F606W$ (broadband
$V$) and $F814W$ (broadband $I$)  filters. The cosmic ray cleaned
images  were obtained from the STScI data archive and have been
processed with the standard ACS pipeline. The images of the two
galaxies are shown in Fig.~\ref{fig:ima_spa} (combined
distortion-corrected  mosaic images corresponding to an exposure time
of 8600s). 

The photometry of resolved stars in the galaxies was made with the ACS module 
of the DOLPHOT package\footnote{\url{http://purcell.as.arizona.edu/dolphot/}} 
for crowded field photometry \citep{dolphin02} using 
the recommended recipe and parameters. Charge transfer efficiency 
corrections were made according to ACS ISR03-09, and the transformation
into the standard $UBVRI$ photometric system were made according to 
\citet{sirianni05}. We 
estimate the uncertainties in the calibration to be 0.05 to 0.10 magnitudes.
Note, however, that 
we used these standard magnitudes for reference only; all our measurements 
were made in the original ACS photometric system to avoid additional 
systematic uncertainties.

\begin{figure*}
\includegraphics[height=12cm]{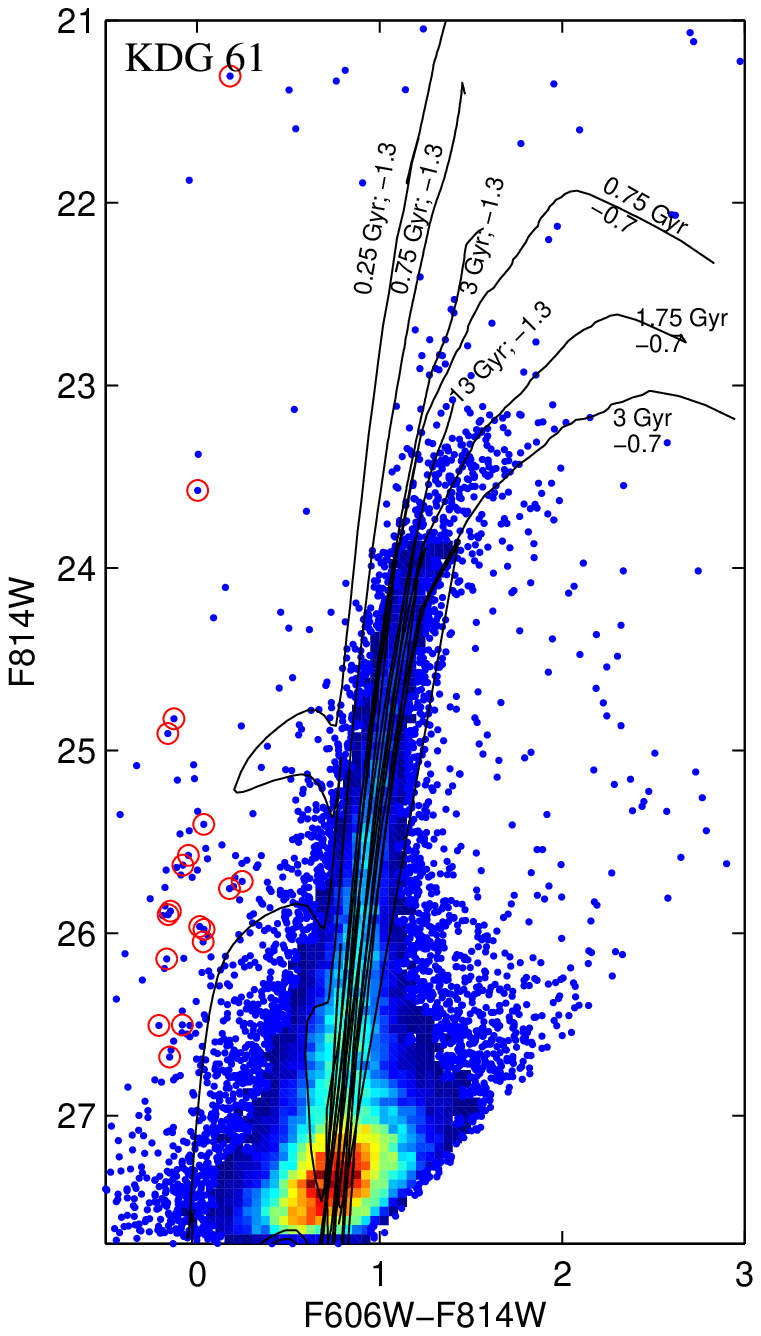}
\includegraphics[height=12cm]{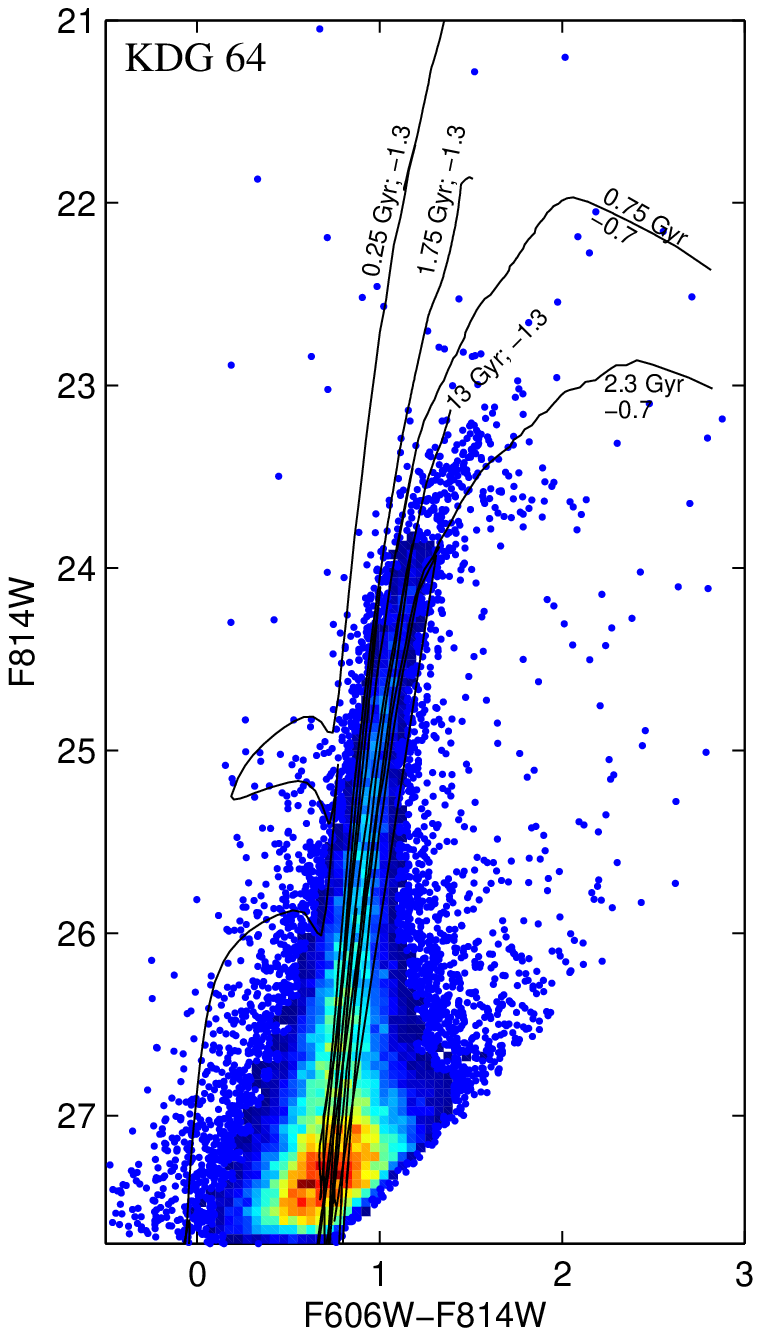}
\caption{The $(F606W-F814W)$, $F814W$ CMDs of the dwarf 
galaxies KDG\,61 (left) and KDG\,64 (right). In the dense parts of the
diagram the colour codes the density in the Hess diagram, while individual stars
are represented where they can be individually distinguished.
The Padova theoterical isochrones \citep{girardi00} corresponding to the mean 
age and metallicity of detected star formation episodes were overplotted.
The isochrones are labelled with the age in Gyr and with the [Fe/H] metallicity.
The magnitudes are not corrected for Galactic extinction.
Blue stars resolved in the \hii{} region of
KDG\,61 are marked with open circles.}
\label{fig:cmd}
\end{figure*}

Only the stars with photometry of good quality were included in the 
final catalogue, following the recommendations given in the DOLPHOT User's 
Guide. We have selected stars with signal-to-noise (S/N) of at least 
five in both filters, 
$\chi^2\le5.0$ and $\vert sharp \vert \le 0.3$.
The resulting CMDs of the two dSphs are presented 
in Fig.~\ref{fig:cmd}. The CMD of KDG\,61 contains 59554 stars and the
one of KDG\,64 contains 45779 stars.

\begin{figure*}
\includegraphics[width=8cm]{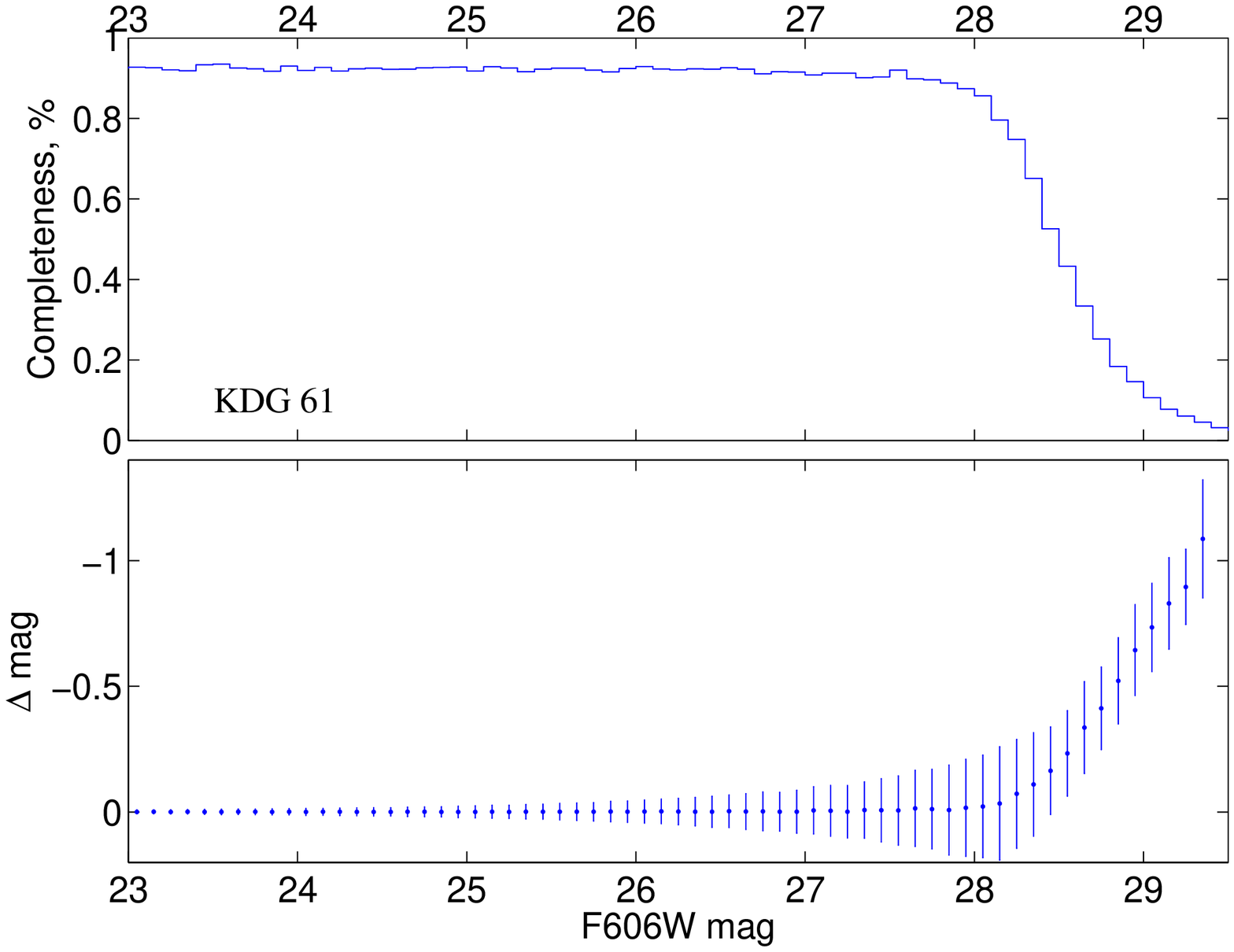}
\includegraphics[width=8cm]{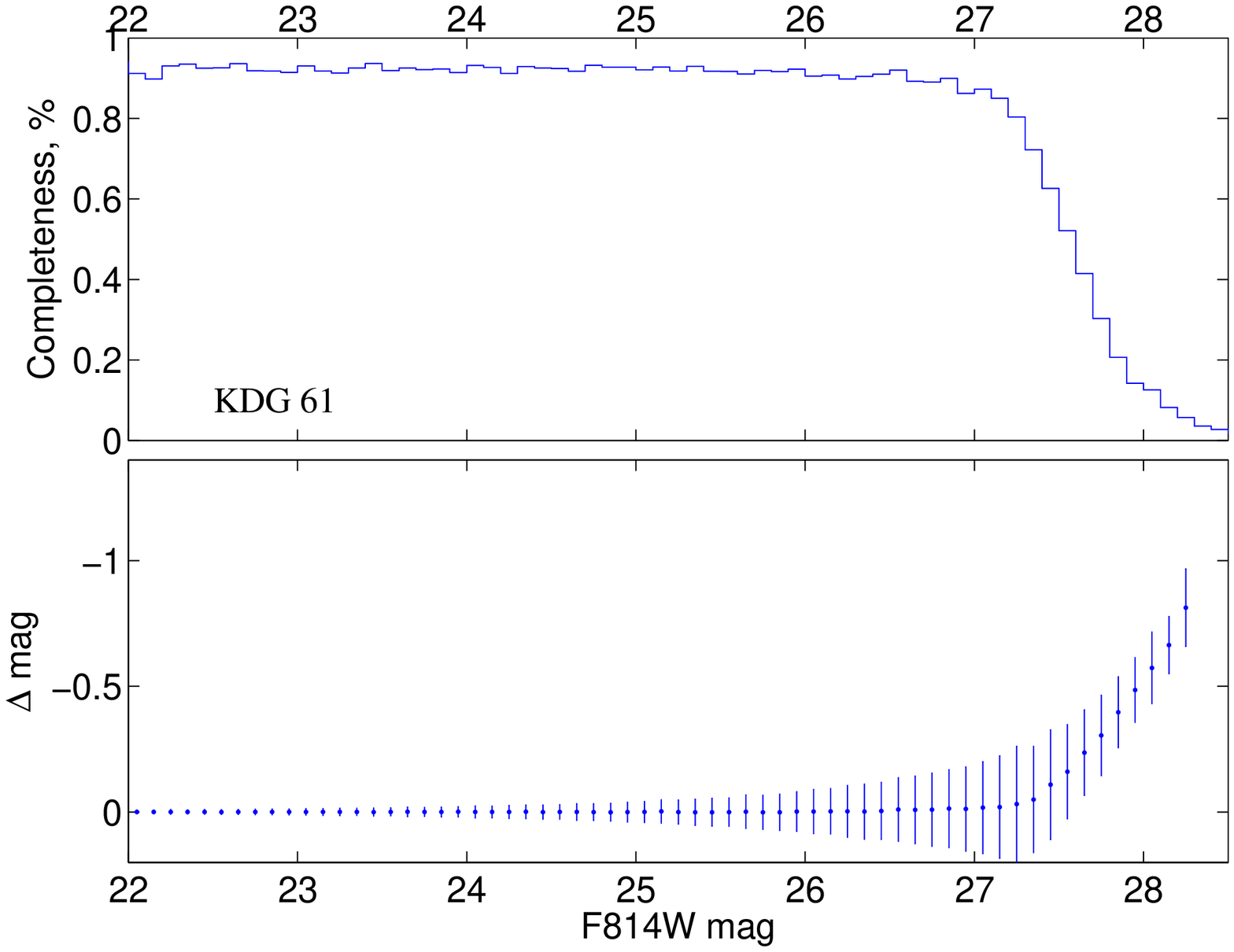}
\caption{Photometric errors and completeness for KDG\,61.
The top panels show the completeness, i.e.\ the fraction of artificial stars
recovered within the photometric reduction procedure, as a function of 
the $F606W$ and $F814W$ magnitudes.
The bottom panels give the difference between the measured and the true input
magnitude ($\Delta$mag = measure $-$ input). The error bars are 1\,$\sigma$
residuals.}
\label{fig:uncert}
\end{figure*}

Artificial star tests is the only accurate way 
to solve the problems of photometric errors, blending and 
incompleteness (see, for example, \citealt{gallart}, 
\citealt{dolphin02}).
These tests were performed for both galaxies using the same
reduction procedures. A large library of artificial stars 
was generated 
spanning the necessary range of stellar magnitudes and colours so that
the distribution of the recovered photometry is adequately sampled. 
The photometric errors and completeness, represented in Fig.~\ref{fig:uncert},
are similar for the both galaxies. 
The 1\,$\sigma$ photometric precision is about 0.08\,mag at  $F814W=26$ and  
0.18\,mag at $F814W=27$. The Malmquist bias becomes notable for stars 
with $F814W > 27.2$. In $F606W$ the 1\,$\sigma$ photometric precision 
is about 0.05\,mag at 26\,mag and about 0.21 at 28\,mag.

\subsection{Colour-magnitude diagrams}
The CMDs of both KDG\,61 and KDG\,64 (Fig.~\ref{fig:cmd})
are typical for 
dSphs, mostly populated with old red giant branch (RGB) stars. 
The red giant branch of KDG\,61 is situated at the magnitude $F814W\,\ge\,23.8$.
The mean colour of the RGB is $(F606W-F814W)\,\simeq\,1.1$ at the absolute 
magnitude level $M_I = -3.5$ according to our measurements. The colour
spread in the red giant branch is not large for this galaxy, 
indicating little or no scatter in metallicity.
Other prominent feature in the CMD of KDG\,61
can be seen above the RGB in magnitude range 
$22.5\,\le\,F814W\,\le\,23.8$. These stars likely represent
an intermediate age asymptotic giant branch population.
There are probably a small number (about 40) of young main sequence blue stars
in KDG\,61 with a mean colour $(F606W-F814W)\,\sim\,0.0$, 
and $23.2\,\le\,F814W\,\le\,26.6$. 
There are also 6 brighter
bluish stars in magnitude range of $21\,\le\,F814W\,\le\,22$ and 
colour range of $0\,\le(F606W-F814W)\,\le\,1$. The spatial distribution 
of the blue stars will be considered in more details below.

The colour-magnitude diagram of the dSph galaxy KDG\,64
is similar to the one of KDG\,61. We can 
see RGB and intermediate age AGB stars in the same range of magnitudes
and colours as in KDG\,61. However, there are no signs of young blue
main sequence at this CMD. 

The deep, high-precision photometry of the dSph galaxies allows us to
estimate the mean metallicity of the red giant branch using the formula of
\citet{lee93} : ${\rm [Fe/H]} = -12.64+12.6(V-I)_{-3.5}-3.3(V_I)_{-3.5}^2$, 
where 
$(V-I)_{-3.5}$ is the mean RGB colour at the absolute magnitude $M_I = -3.5$.
According to this equation the mean $\textrm{[Fe/H]}$ ratio is equal to 
$-1.49\,\pm\,0.03$\,dex for KDG\,61 and $-1.59\,\pm\,0.02$\,dex for KDG\,64.

\subsubsection{Foreground contamination}
The CMDs of both KDG\,61 and KDG\,64 are evidently 
not heavily 
contaminated by foreground stars, but as we will perform a detailed analysis of
the star formation history, we will make a quantitative assessment of this
pollution.

We could not find an appropriate field observed with the HST nearby to 
the galaxies under study, which would have given us the possibility to estimate 
a number of foreground stars in these CMDs. 
An alternative, in this case, is to use simulated
star counts in our Galaxy. We have constructed the simulated CMDs of the 
Galactic stars in the regions of KDG\,61 and KDG\,64 using 
\textsc{TRILEGAL} \citep{girardi05}. 
\textsc{TRILEGAL} simulates CMDs in the ACS instrumental system taking into
account the components of thin and thick Galactic disks, the halo and the bulge
of our Galaxy. The photometric errors, saturation and incompleteness 
were taken into account in the modelling. These models confirm
the negligible contamination by foreground stars. 
Thus, the expected number of foreground stars in our CMDs would be about 
40 up to the photometric limit. 
We have used also the Besancon model of stellar population synthesis of
the Galaxy \citep{robin} to estimate a number of foreground stars in the fields
of KDG\,61 and KDG\,64. The Besancon model accounts for the four galactic
populations: thin and thick disk, spheroid and bulge. The photometric errors were 
applied to the modelling. The resulting star counts give about 30 stars in
our object fields. The synthetic Milky Way stars, according to the TRILEGAL 
simulations, falls in the magnitude range $20 < F814W < 28$ and 
the colour range $0.8 < F606W-F814W < 2.5$. 

In our previous work \citep{makarova02} we already estimated the number of 
the Milky Way stars in the HST/WFPC2 images of dwarf galaxies in the area
of M\,81 group. We have analysed observed colour-magnitude diagrams of 24 dwarf
galaxies and also theoretical star counts of \citet{bahcalandsoneira}. 
According to our measurements, we expect between 1 to 5 foreground stars 
in the magnitude range $18\fm0 \le I \le 22\fm0$ and in the colour
range $1.4 \le (V-I)_0 \le 3.0$ in WFPC2 field. This is in a good
agreement with the present \textsc{TRILEGAL} simulations, which give 
us about 4 stars in the same magnitude and colour range, taking into 
account the field difference between ACS and WFPC2.

\subsubsection{Spatial distribution of the blue stars}
\label{sect:blue}
The scarce population of blue stars described above may trace an
young star formation event. Therefore, we examine closely their location.
We selected blue stars with $(V-I)_{0} < 0.4$\,mag and $I_0 < 26.5$\,mag. 
In KDG\,61, 2/3 of the 45 selected blue objects are randomly 
distributed over the field of the ACS image. 
The one-sample Kolmogorov-Smirnov test has shown the homogeneous
distribution of the stars. Moreover, the distribution of the blue stars in
the ACS field is significantly different from the distribution of all other (red)
stars. The two-sample Kolmogorov-Smirnov test rejects the hypothesis of the same
distribution of the red and blue stars at the 0.006 significance level, what leads
us to the conclusion that the blue stars do not belong to KDG\,61 itself.
Probably, the bluish colour of most of the detected sources is due to 
a photometric contamination by a bright close companion. 
The rest is associated to the \hii{} region in KDG 61 and to a small and 
loose concentration of the bluish stars at the North-West of the galaxy's body.
This latter concentration contains only 7 blue stars and we have not enough data 
to discuss the nature of group of stars associated with it.
The two regions are indicated with circles in Fig.~\ref{fig:ima_spa}, 
and the blue sources associated with the \hii{} region are marked in 
Fig.~\ref{fig:cmd}.

There is no pronounced blue star population in KDG\,64. Dozen of bluish stars,
situated outside RGB in the colour-magnitude diagram, are spread
over the ACS image and most of them have brighter close companion.

\begin{figure*}
\includegraphics[width=8cm]{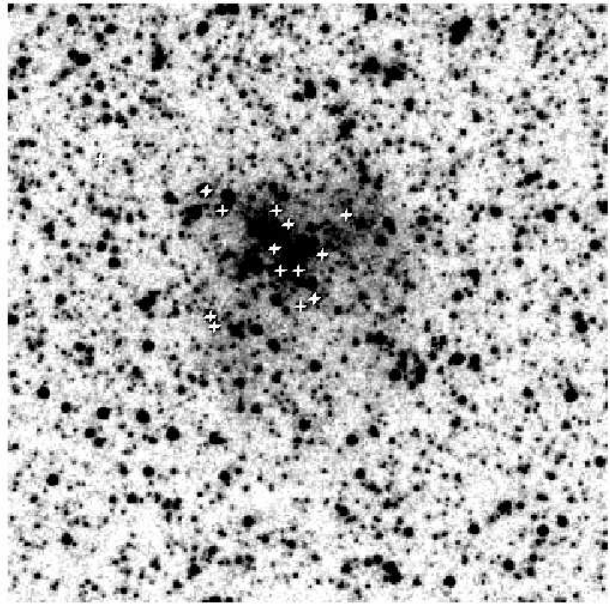}
\caption{$F814W$-band image of \hii{} region in KDG\,61. 
The picture size is 15\,arcsec (250\,pc). We mark with crosses 
the resolved blue stars at this region.}
\label{fig:hii_reg}
\end{figure*}

\subsection{Distance determination}
The precise knowledge of the distance is crucial to determine
the star formation history from CMD analysis. 
The photometric tip of the red giant branch (TRGB) distances to the galaxies 
were previously obtained by \citet{k00} using HST/WFPC2 images. 
The true distance moduli given in that work are $27.78\pm0.15$ for KDG\,61 
and $27.84\pm0.15$ for KDG\,64. The corresponding distances are 3.6 Mpc 
and 3.7 Mpc, respectively. \citet{recola05} has obtained 
distance moduli of $27.61\pm0.17$ ($D = 3.3$ Mpc) for KDG\,61 and 
$27.74\pm0.18$ (3.5 Mpc) for KDG\,64  using surface brightness fluctuation (SBF) 
method based on $B$- and $R$-band images obtained with the Nordic Optical 
Telescope. 
Both measurements give consistent distances for our galaxies although 
the SBF measurements seem to give a little shorter scale. 
However, the difference is inside the error bars.

The present photometry is deeper and it has a higher quality, due to the 
advantage of the ACS detectors in comparison to the WFPC2. There are also
a number of recent improvements implemented to the TRGB method itself. 
We have determined the photometric TRGB distances with
our {\it trgbtool} program, which uses a maximum-likelihood algorithm 
to determine the magnitude of tip of the red giant branch from the stellar
luminosity function \citep{makarov06}. 
The measured TRGB magnitudes are $F814W_{TRGB} = 23.84\pm0.02$ for KDG\,61 
and $F814W_{TRGB} = 23.87\pm0.02$ for KDG\,64 in the ACS instrumental system.
As pointed in \citet{rizzietal07}, reaching measurements of this quality, 
the precision on the Galactic extinction
becomes a major source of uncertainty. The precision of the extinction on
the Schlegel's maps is 16 per\,cents, resulting in an uncertainty of 0.02\,mag in the
$I$ band. The last source of error is the calibration of the TRGB distance
indicator.
Using the calibration for the TRGB distance indicator by 
\citet{rizzietal07}, and adding all the sources or errors, 
we derived the true distance moduli for 
KDG\,61: $\mu(0) = 27.77\pm0.04$ ($D = 3.58\pm0.07$ Mpc) and KDG\,64: 
$\mu(0) = 27.84\pm0.04$ ($D = 3.70\pm0.07$\,Mpc).
These new distances are in a good agreement with the previous estimations and
they have a better precision.

TRGB distances for several galaxies in M\,81 group were recently measured also by
\citet{dalcanton09} within their ANGST project.
The distance moduli of KDG\,61 (27.72) and KDG\,64 (27.85) are in good 
agreement with our values.

The original Cepheid distance to  M\,81, $3.63 \pm 0.34$~Mpc \citep{freedman94},
was revised to $3.55 \pm 0.13$ Mpc \citep{freedman01} ($\mu = 27.75 \pm 0.08$).
This value is in close agreement with the TRGB distance from \citet{dalcanton09}.
Recently an independent geometric estimation using the size of the expanding 
shell around SN1993J gave the distance $3.96 \pm 0.29$~Mpc \citep{bartel07}.
It is also consistent within the uncertainties, but larger than the two
other determinations by 10 per\,cents. Discarding this last value,
we adopt a distance $3.58 \pm 0.04$~Mpc.
Beside the value from \citet{dalcanton09}, the distance to NGC\,3077 was 
determined by the surface brightness fluctuation  \citep{tonry01}
$\mu(0) = 28.03 \pm 0.13 $ ($4.04 \pm 0.15$ Mpc)
and TRGB methods \citep{karachentsev03}
 $\mu(0) = 27.91$ ($3.82$ Mpc). The two TRGB distances agree, and
the SBF measurements is larger. We adopt the ANGST value
$\mu(0) = 27.92 \pm 0.02 $ ($3.83 \pm 0.04$ Mpc)

These accurate distances for the two dwarf galaxies allow us to
estimate their location within the group of galaxies, and more particularly
with respect to M\,81 and NGC\,3077.
Taking into account
the distances and angular separation of the objects on the sky,
we estimate the spatial separation to M\,81
to be about 40\,kpc for  KDG\,61 and about 230\,kpc 
for  KDG\,64.
KDG\,61 is situated very close to the central body of the group.
The only galaxies which are likely closer to M\,81 are 
Holmberg IX and BK3N tidal dwarfs.
KDG\,64 is probably situated on the back side of M\,81 and possibly 
slightly in front of NGC\,3077. The spatial separation with NGC\,3077
is about 155\,kpc.
Both KDG\,61 and 64 are located at the same distance to M\,81 as the
dSph satellites of the Milky Way (60\,kpc to 440\,kpc).

\subsection{The star formation histories}
\label{sect:sfh}
The star formation and metal enrichment histories 
of KDG\,61 and KDG\,64 were determined from 
their CMDs using our StarProbe package \citep{mm04}. 
This program adjusts the observed photometric distribution of stars in the 
colour-magnitude diagrams against a positive linear combination 
of synthetic diagrams of single stellar populations (SSPs, single age
and single metallicity). Our approach is similar to the technique described 
in the works of \citet{tosi1989}, \citet{aparicio1997} and
\citet{dolphin2000}.

The observed data were binned into Hess diagrams, giving the
number of stars in cells of the CMDs (two-dimensional histograms).
The size of the cells are 0.07\,mag in luminosity and colour.
They are large enough to contain a significant number of stars,
and small enough to trace the characteristic features of the CMD distribution.
The synthetic Hess diagrams were constructed from theoretical stellar 
isochrones and initial mass function (IMF). 
Each isochrone describes the magnitudes and colours of a stellar population 
with a particular age and metallicity as a map of probabilities
to find a star in each cell.
We used the Padova2000 set of theoretical isochrones \citep{girardi00}, and a 
\citet{salpeter55} IMF. The distance is adopted from the present paper 
(see above) and the Galactic extinction from \citet{schlegel98}.
The synthetic diagrams were altered by the same
incompleteness and crowding effects, and photometric systematics
as those determined for the observations using artificial stars 
experiments.

We have taken into account the presence of unresolved binary stars
(binary fraction). 
As there is no simple and well accepted solution to describe 
the co-evolution of the close interacting binaries, 
we used the standard evolution models.
The effect of interacting binaries in population models has been found 
to be small \citep{zhang2009}.
Following \citet{barmina}, the binary fraction was taken to be 30 per\,cents. 
The mass function of individual stars and the main component 
of a binary system is supposed to be the same. 
The mass distribution for second component was taken to be flat 
in the range 0.7 to 1.0 of the main component mass.

Besides unresolved binary stars a photometric blending (detection of several 
stars as a single source) should be presented. 
A blending is sufficient in crowded fields like in globular clusters and 
in central regions of galaxies.
Occasional superposition of stars should distort observational stellar profile.
That stars are later excluded by DOLPHOT parameters like sharpness, roundness and 
characteristic of the PSF-fitting. 
This effect is taken into account in completeness function constructed from
artificial star photometry.
However, very close optical binary stars could not be excluded in this way.
We have estimated a blending probability for stars brighter than 28.0 mag 
in $F814W$ band.
For both galaxies it is not exceeded $2\cdot10^{-4}$~per\,cent.
Therefore, tacking into account the amount of detected stars the expected number 
of blended stars is 0.1 over the ACS field. Thus the effect is negligible.

The basis of synthetic diagrams covers
all the range of ages (from 0\,Myr to 14\,Gyr) and metallicities 
(from $Z = 0.0001$ to $Z = 0.03$), with an age resolution which
increases for young populations (2 Gyr steps for the old populations; 
0.5 Gyr near 1 Gyr and younger), and with a metallicity resolution
limited by the underlying published isochrones. The isochrones were
interpolated in age, to avoid discontinuities, so that the sampled points 
in the CMD are separated by at maximum 0.03 mag. They
were not interpolated in metallicity.

\begin{figure*}
\includegraphics[width=8cm]{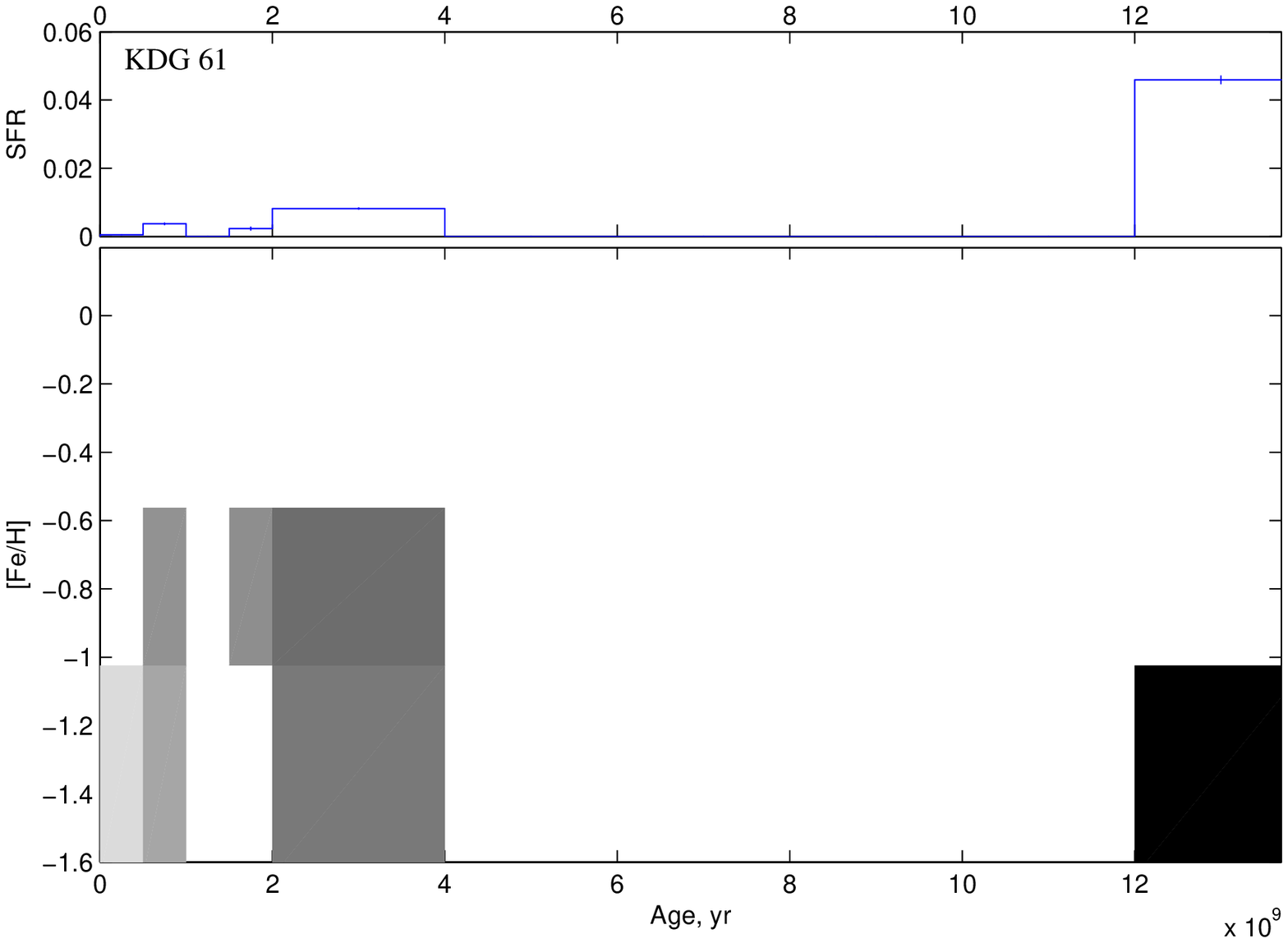}
\includegraphics[width=8cm]{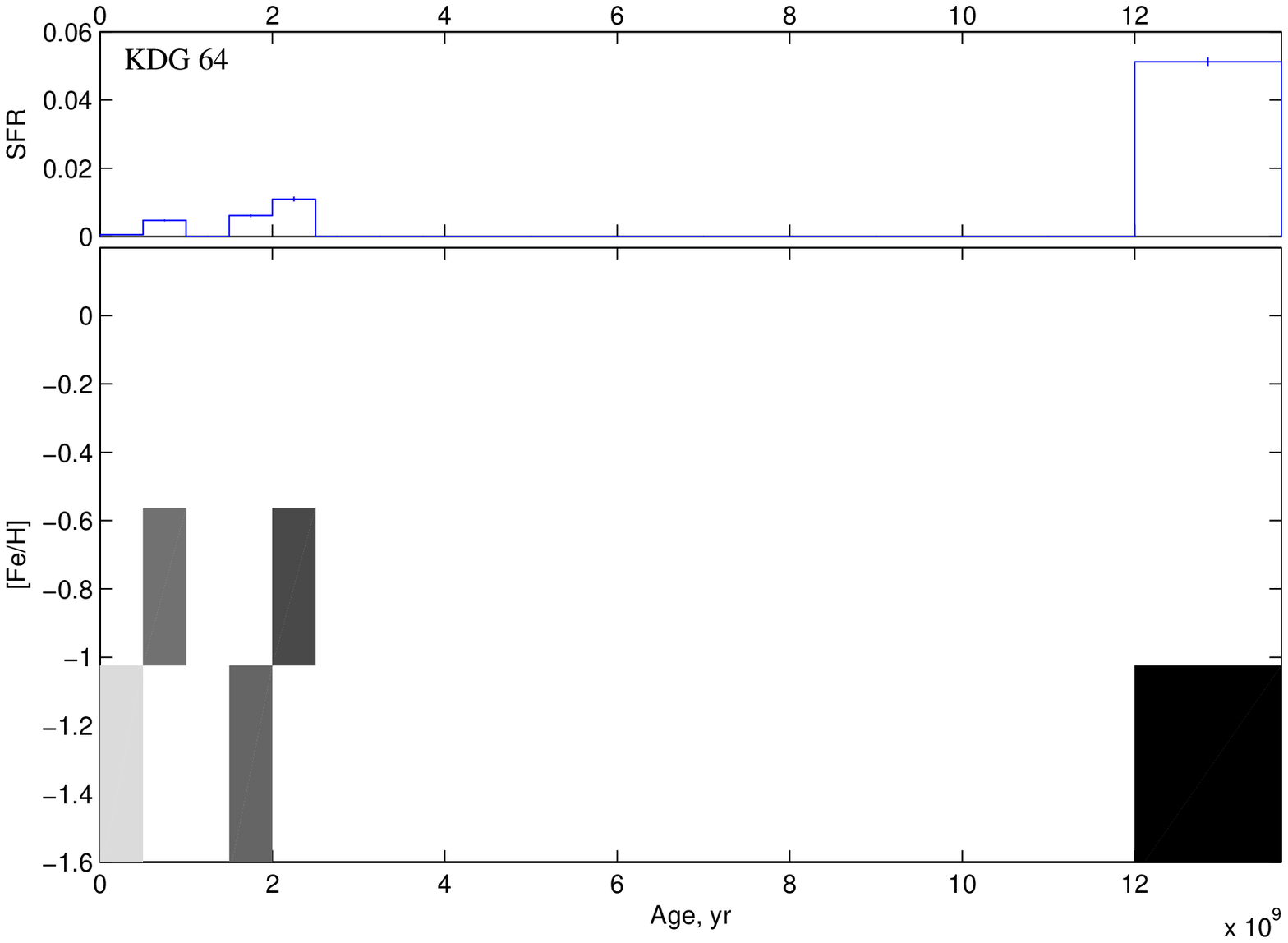}
\caption{The star formation histories of the dwarf galaxies 
KDG\,61 (left) and KDG\,64 (right).
Top panels show the star formation rate (SFR) ($M_\odot$/yr) against the
age of the stellar populations.
The bottom panels represent the metallicity of stellar content as function of 
age. The grayscale colour corresponds to the strength of SFR for given age and 
metallicity.}
\label{fig:cmd_sfh}
\end{figure*}

The best fitting combination of synthetic CMDs is a maximum-likelihood solution 
taking into account the Poisson noise of star counts in the cells of Hess diagram.
The resulting star formation history (SFH) are shown in 
the Fig.~\ref{fig:cmd_sfh}.
The 1\,$\sigma$ error of each SSP is derived from analysis of likelihood function.

For both galaxies the main star formation event occurred
between 12 and 14\,Gyr. These initial bursts account for 82 and 89 per\,cents 
of the total mass of formed stars for KDG\,61 and KDG\,64 respectively.
The range of metallicity is about ${\rm [Fe/H]}=[-1.6:-1]$\,dex. 
The mean SFR is high in this period for the both objects: 
$4.5\,10^{-2}$ \msol{} yr$^{-1}$ for KDG\,61 and 
about $5.2\,10^{-2}$ \msol{} yr$^{-1}$ for KDG\,64. 
These are total star formation rates over the whole galaxy in the ACS image. 
A rescaling to the total mass of the galaxy would increase the rates by 
a few percent for KDG\,64.

There are indications of intermediate age star 
formation in KDG\,61 between 2 and 4\,Gyr ago and of more recent star 
formation about  0.5--1\,Gyr ago. 
The measured metallicity range is higher for this period: 
${\rm [Fe/H]}=[-1.6:-0.6]$ dex, indicating possible slight metal enrichment
for the younger star formation episodes.

\begin{figure*}
\includegraphics[width=8cm]{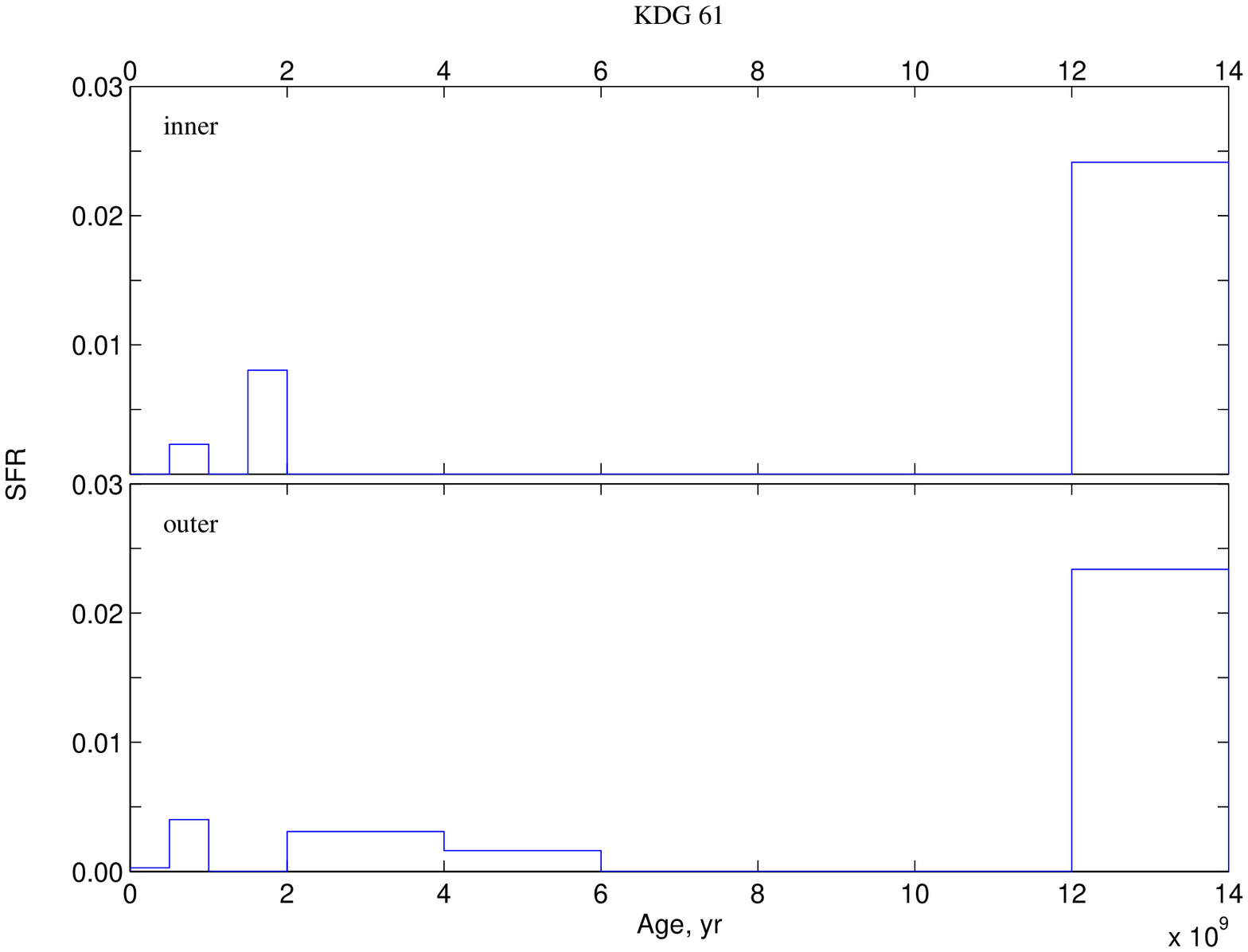}
\includegraphics[width=8cm]{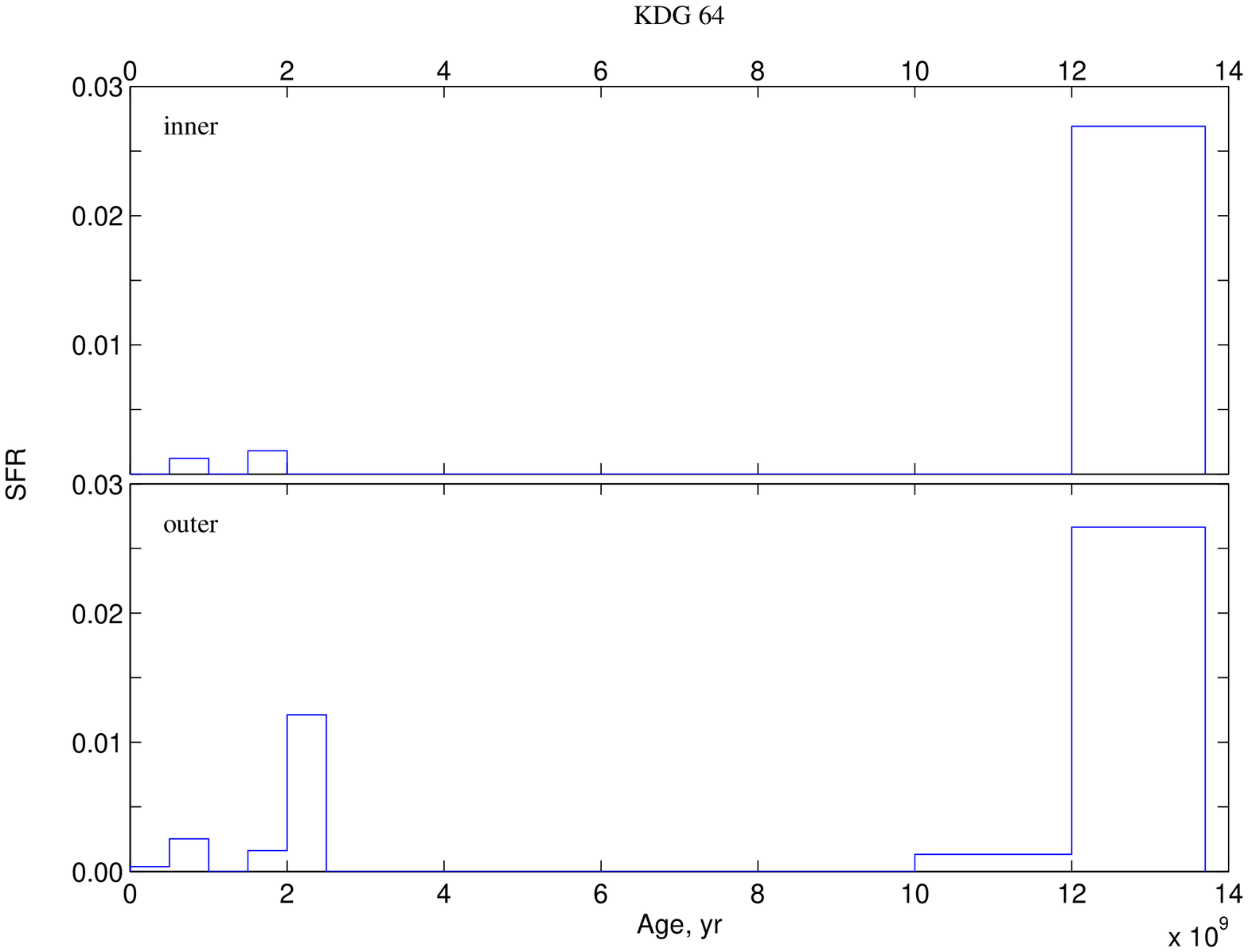}
\caption{The star formation histories of the dwarf galaxies 
KDG\,61 (left) and KDG\,64 (right) in the inner and outer regions.
All panels show the star formation rate (SFR) ($M_\odot$/yr) against the
age of the stellar populations.}
\label{fig:sfh_inout}
\end{figure*}

We have measured star formation history in two separated regions.
For the first sample the stars located within 30 arcsec 
from the centre of KDG\,61 were selected, and for the second sample 
the outer stars were selected. 
While the older dominant population is homogeneous, the intermediate
age population appears younger (1.5--2 Gyr) in the inner region than
in the outskirts (2--4 Gyr). 
This result is illustrated in the Fig.~\ref{fig:sfh_inout} (two left panels).
At the upper panel the SFH in the inner region is shown, and in the lower panel 
the outer region is shown.

In the case of KDG\,64 
we detect a slight enhancement of the star formation about 1.5--2.5\,Gyr ago
and a
very small fraction of younger population can have an age of about 500 Myr.
A possible metal enrichment over the galaxy lifetime, 
similar to KDG\,61, can be present. 
As for the previous galaxy, we have measured the star formation 
history of the two separated samples of the resolved stars (inside and outside
30 arcsec). A marginal sign of spatial 
separation of the stars with different
ages was found: The stars of 1.5--2 Gyr dominate in the inner part and
stars of 2--2.5 Gyr in the outer part (see Fig.~\ref{fig:sfh_inout}). 
As in KDG\,61, the older stars are homogeneously distributed.
The metallicity of the star formation episodes did not show sufficient differences
in the inner regions in comparison to the outer region in the both galaxies.

\section{Integrated light spectra}
We use the integrated light spectra of KDG\,64 and KDG\,61 to 
determine the characteristics of the stellar population.
The star formation history reconstruction is affected by 
the well known age-metallicity degeneracy. This is true for 
colour-magnitude diagrams as well as for integrated light spectra. 
Simultaneous usage of the both methods, for the galaxies under study, gives 
us a hope to improve the reliability of age and metallicity determination.

\subsection{Observations and data reduction}
We obtained spectra of KDG\,64 with the SCORPIO 
\citep{scorpio} focal reducer of the 6-meter telescope of the 
Special Astrophysical Observatory of the Russian Academy of Sciences.
The observations were performed in long-slit mode during dark time with 
2k$\times$2k CCD EEV42--40 detector
(2005 Oct. 6th; PI Philippe Prugniel; Observers Victor Afanasiev 
and Olga Sil'chenko). We used the 2300 lines per mm grism, which 
together with a slit width of 1\,arcsec provides a spectral resolution of
2.2\,\AA{} (FWHM, corresponding to an instrumental velocity 
dispersion $\sigma_{\rm ins}\,\simeq\,56$\,\kms{} or $R\,\approx\,2300$), 
a dispersion of 0.38\,\AA/pix and a spectral range of 4700--5500\,\AA. 
The frames were binned by 2 pixels along the slit direction, resulting
in a spatial scale of 0.357\,arcsec/pix.
The observations were made along the position angle, ${\rm PA}=24^\circ$, 
corresponding to the major axis of the galaxy.
We obtained 9 spectra, each of 20 minutes, resulting in 3 hours in total. 
We performed arc-lamp (He-Ne-Ar) calibration at every hour, as well as 
in the beginning and the end of the night. Thus, we improve the 
quality of the wavelength calibration, accounting for possible drift 
due to flexures. Twilight spectra were taken at the beginnings of each 
night. We used them to correct the variation of the illumination 
along the slit and to determine the line-spread function of the system.
The seeing during these observations was 4.1--4.6\,arcsec.

KDG\,61 was observed on 2007 April 17th and 21st with the same 
observational set-up. The spectra taken during the first night were 
not used, because of a low quality. Thus, we were left with
six exposures of 1200~sec (2 hours). 
The slit was oriented along the major axis (${\rm PA}=54^\circ$). It
corresponds to the line joining the central cluster with the \hii{} region.

We attempted to use the spectra obtained by \citet[2001 on January 18 
and 23][]{sharina01} with the UAGS long-slit spectrograph at the same 
telescope. Despite comparable exposure times, these spectra have lower S/N
due to the less efficient instrument and detector and are not useful to 
complete our data. 

The reduction was made in the ESO-MIDAS package with the \textsc{LONG}
context. It includes bias subtraction, flat field correction and wavelength 
calibration. 
The most critical part of the data reduction is the sky subtraction, as
the brightness in the centre of these galaxies is of the order of 
10--20\,per\,cents of the brightness of the sky.
The main difficulty is the change of the line-spread function 
(or spectral PSF; hereafter LSF) with the position along the slit. 
The LSF is the
imprint of  a spectrograph resolution plus the data reduction
(imperfect wavelength calibration). 
A simple interpolation between two sky regions as proposed in MIDAS is
not appropriate because the LSF
is narrower in the centre where the galaxy is placed than in the edges
where the sky can be estimated. This pattern is however not symmetrical,
and we extracted the sky from a region of the slit were the LSF is 
almost similar to the one in the galaxy region. We determined the
relative LSF between the sky and object extractions \citep[see][]{ulyss} and
injected it in the galaxy spectrum before subtracting the sky.

There is a significant H$_\beta$ emission line in the sky. It is partly due to
the geocoronal fluorescence emission of the atomic Hydrogen from the 
thermosphere and exosphere ionized by the solar radiation 
at altitudes between 500 and 80000 km \citep{sahan07} and partly due to a 
possible Galactic contribution. As the redshift of our targets are nearly zero,
the residuals of the subtraction of this component can alter the centre 
of H$_\beta$.

\subsection{Analysis method}
We use ULySS\footnote{\url{http://ulyss.univ-lyon1.fr}}  \citep{ulyss}
to analyse the long-slit spectra. This
program minimizes the \csq{} between the observations and  a
combination of SSP models, to fit the characteristics
of  the population and the line-of-sight velocity distribution
(LOSVD) at the same time. 
The program uses all
the pixels of the spectrum and optimally exploits the available information.
It fits the observation data as
\begin{eqnarray}
{\rm{ Obs}(\lambda)} = P_{n}(\lambda) &\times& 
    LOSVD(v_{sys},\sigma) \otimes \nonumber \\
    & & \sum_{i=0}^{i=m} W_i \, SSP_i({\rm Age}, {\rm [Fe/H]},\lambda).
  \label{eqn:main}
\end{eqnarray}
The model used in this paper may be either a single SSP or a combination
of several SSPs parametrized by their age and metallicity ($\textrm{[Fe/H]}$). 
$W_i$ are the weights of each of the SSPs. The SSPs are convolved by the LOSVD
which is parametrized by the systemic velocity ($v_{sys}$) 
and the velocity dispersion ($\sigma$). A multiplicative polynomial of 
order $n$, $ P_{n}(\lambda)$, renders the method insensitive to the 
extinction an imperfectness in the flux calibration. For this paper 
we find $n=20$ to be the optimal order of the polynomial.
More details on the analysis method can be found in \citet{ulyss}.
The retrieved solution and errors are checked via \csq-convergence map and
Monte Carlo simulations. 
The full spectrum fitting was extensively validated with Galactic
globular clusters in \citet{koleva08a} and was used to study dE
galaxies in \citet{kol2009}.

Our SSPs spectra are computed with Pegase.HR code \citep{pegasehr}
using Padova1994 isochrones \citep[][and companion papers]{padova94}, 
Salpeter's IMF \citep{salpeter55} and Elodie.3.1 stellar library 
\citep{elodie1, elodie31}. 

\subsection{KDG\,61}
\label{subs:k61}

\begin{figure}
  \includegraphics{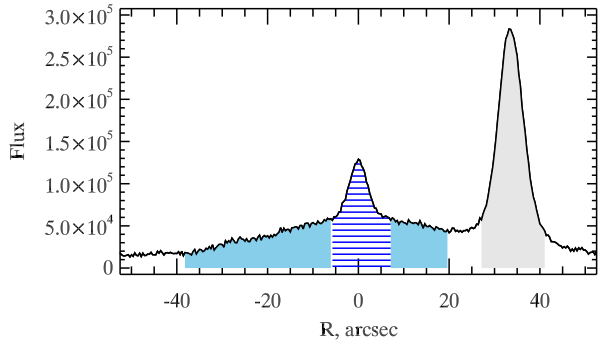}
  \caption{Luminosity profile of KDG\,61. The grey shadowed area, centred
    at about 35\,arcsec, shows
    the extracted region representing the \hii{} knot. The blue shadowed area
    is the galaxy's region. The horizontal blue lines mark the 
    globular cluster region.}
  \label{fig:k61_profile}
\end{figure}

We could not perform 2D analysis and investigate the stellar 
population parameters along the galaxy's radius due to low S/N. 
However, we could
reach $\textrm{S/N}\ge5$ binning the spectra in three different regions. 
The first one is the central part of 14\,arcsec containing the globular 
cluster. The second is the \hii{} knot. The third is  the 
galaxy itself excluding the previous regions and extended to a diameter 
of 40\,arcsec.  We did optimal extraction weighting the individual
spectra by their errors. The location of these three extractions is presented in 
Fig.\ref{fig:k61_profile}. This three extractions were analysed separately.

The next step of the analysis is to determine precisely the LSF.
We modelled it with a Gaussian plus Gauss-Hermite distortion terms 
changing with the wavelength and position along the slit.  The LSF 
was injected in the models \citep[see][]{ulyss} to take into account 
the variation of the resolution with the wavelength and to correct 
the small residual errors in the wavelength calibration.

We use the twilight sky to determine the wavelength dependence of the LSF.
We also analysed the background sky of the galaxy against high resolution
dark sky spectra obtained with UVES \citep{hanuschik03}
convolved by 13 \kms{} (to match the resolution of our models).
We used the dark sky between 4800\,\AA{} to 5000\,\AA{} 
to adjust the zero-point of the velocity. These values were consistent 
between the globular cluster's and galaxy's extractions. 
In this paper we do not present or discuss  the derived physical 
velocity dispersion, $\sigma$, because it is highly uncertain.

\subsubsection{\hii{} knot}
We fitted the position, the width and the intensity of [O{\sc iii}]4958, 
[O{\sc iii}]5007 and  H$_{\beta}$ using {\sc uly\_line}. 
For the three lines we have heliocentric corrected cz of
$-126$, $-126$ and $-117$\,\kms{} with internal precision of 3\,\kms{}
and systematic uncertainty of  5\,\kms{} (arising from the determination of
the velocity zero-point). We adopt an average ${\rm cz}\,=\,-123\pm6$\,
\kms{} which is in agreement with the two previous measurements
$-116\pm21$\,\kms{} \citep{sharina01} and $-135\pm30$\,\kms{}
\citep{johnson97}.

\subsubsection{Central globular cluster}
The spectrum of the globular cluster was fitted against a grid of SSPs. 
We obtained heliocentric velocity of ${\rm cz}\,=\,222\pm3$\,kms{}.
This velocity is in qualitative agreement with the result from 
\citet{croxall09}, showing the H$_\beta$ absorption line of
the cluster redshifted with respect to the \hii{} knot. 
We find that SSP-equivalent age of the globular cluster is $16\pm2$\,Gyr, 
and its metallicity is ${\rm [Fe/H]} = -1.5\pm0.1$\,dex.
The errors were determined using Monte Carlo simulations
(1000 realizations of the noise).

\subsubsection{Galaxy's body}

We have found that the heliocentric velocity of the galaxy is 
${\rm cz}\,=\,221\pm3$\,\kms{}.
We determined a relatively young SSP-equivalent age $1.7\pm0.5$\,Gyr
and ${\rm [Fe/H]}\,=\,-0.9\pm0.2$\,dex. The errors were determined
using 1000 Monte Carlo simulations.

The CMD fit revealed the presence of young stars from 0.5 to 3\,Gyr
superimposed on the dominant 13\,Gyr old population (see Sect.~\ref{sect:sfh}).
The S/N of the spectrum is not sufficient to perform an unconstrained
analysis of the star formation history as was done in \citet{kol2009}.
However, bounding the age distribution we may expect to obtain a
tight constraint on the metallicity of the two epochs of star formation.

\begin{figure}
  \includegraphics{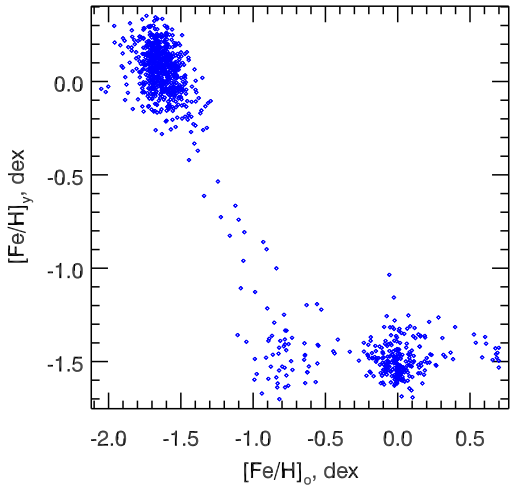}
  \caption{MC simulations for the 2 component fit of K61. In x-axis
    we plot the metallicity of the old component and in the y-axis 
    the metallicity of the young component.}
  \label{fig:k61_mc}
\end{figure}

Therefore, we fitted the spectrum with two SSP models fixing the
ages to those given by the CMD analysis, i.e.\ 13 and 0.8\,Gyr.
It gives us the metallicity and portion of the old component to be
${\rm [Fe/H]} = -1.5 \pm 0.2$ and 86 per\,cents of the stellar mass.
The young population comprises 14\,per\,cents of the mass 
with ${\rm [Fe/H]} =  0.0 \pm 0.3$.
The error bars were determined using Monte Carlo simulations 
(Fig.\,\ref{fig:k61_mc}). It shows
the existence of two solutions. The second one, less populated,
has inverse metallicities between the two bursts and is physically
less plausible.

The metal enrichment of the young with respect to the old population is 
stronger than what is suggested by the CMD analysis. 

\begin{figure*}
  \includegraphics{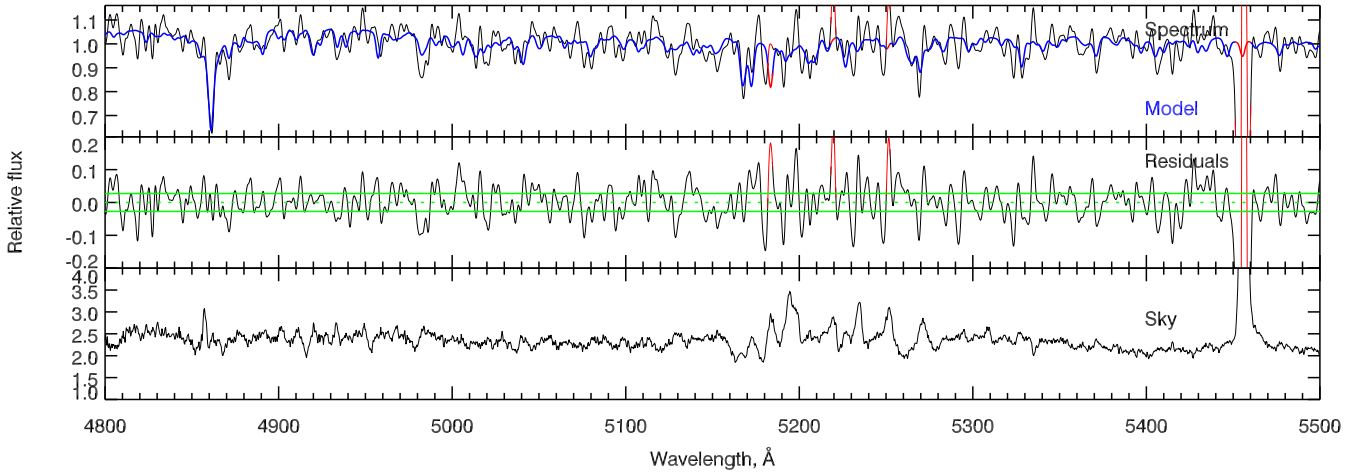}
  \caption{A SSP fit of the central globular cluster of KDG\,61. 
    From top to the bottom:
   with black we plot the spectrum and with blue 
   the best fit model;
   the residuals from the fit (note the extended scale) and the 1\,$\sigma$
   error (in green); the sky 
   extracted from the edge of the 2D image. In all the panels the masked
   pixels are plotted with red.}
\end{figure*}

\subsection{KDG\,64}
\label{subs:k64}

The object in the centre of KDG\,64  \citep{bp94} is known
to be a 1\,arcsec diameter background spiral galaxy 
\citep{k00, sp02, sharina01}.
We have determined the radial velocity of the background galaxy to be
${\rm cz}=57540$\,\kms{} and have fitted its stellar population in order
to properly model its contamination in the KDG\,64 spectrum. 
See the details of the analysis in Appendix~\ref{sect:bg}.

\begin{figure}
\includegraphics{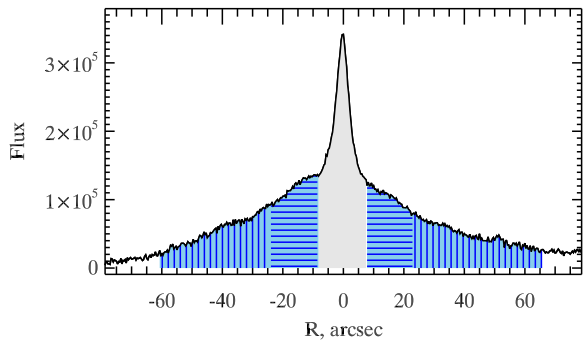}
\caption{Intensity profile of KDG\,64. With grey we mark the extracted region
representing the background galaxy (the central 15\,arcsec). 
With blue we show KDG\,64 extraction and
with dark blue horizontal and vertical lines we show the inner and outer 
extracted regions, respectively.}
\label{fig:k64prof}
\end{figure}

We have extracted the region of the galaxy in a diameter of 110\,arcsec
excluding the 15\,arcsec central region contaminated by the background
galaxy (Fig.\ref{fig:k64prof}). We also split the extraction in
inner and outer regions (as it is shown in Fig.\ref{fig:k64prof}) 
to search for radial changes in the galaxy's stellar population.

As for KDG\,61 (Sect.\ref{subs:k61}) we corrected the zero
point of the velocity using the UVES sky spectrum.

\subsubsection{SSP-equivalent analysis}

The fit against one SSP shows an age of $4\pm2$\,Gyr and metallicity of
$-1.5\pm0.1$\,dex. For the radial velocity we find  
$-15\pm13$\,\kms{} in agreement with $-18\pm14$\,\kms{} from previous 
measurements of \citet{sp02}.

\subsubsection{Star formation history}

For multi-population analysis we tried to
decompose the spectrum in two components: young ($<2.5$\,Gyr) and old
fixed at 10\,Gyr. This choice for the old population gives smaller residuals
than using the age of 13 Gyr inferred from the CMD analysis. We found the young 
component to have an age of $\sim1.6\pm1$\,Gyr which is consistent with 
the CMD studies. However, the mass fraction
in the young component is 38\,per\,cents, which is significantly 
higher than indicated from the CMD analysis (11\,per\,cents).
To determine the values and the errors of the metallicity we performed
1000 Monte-Carlo simulations. The metallicity of the old component is
${\rm [Fe/H]}=-1.6\pm0.4$\,dex. The young stars have almost the same metallicity
${\rm [Fe/H]}=-1.5\pm0.4$\,dex. Therefore the galaxy shows no metallicity 
evolution. This result is consistent with the CMD analysis.

\subsubsection{Radial profiles of the population}

Encouraged from the good results and the higher S/N ($\sim\,8$)
comparable to KDG\,61, we tried to find 
population gradients along the galaxy. 
To be consistent with the 
CMD studies we compared the inner 30\,arcsec
part and the rest of the galaxy.  
We could not find any difference in the two extractions. 
The ages and 
metallicities are the same within the error bars like the general
extraction.

\begin{figure*}
\includegraphics{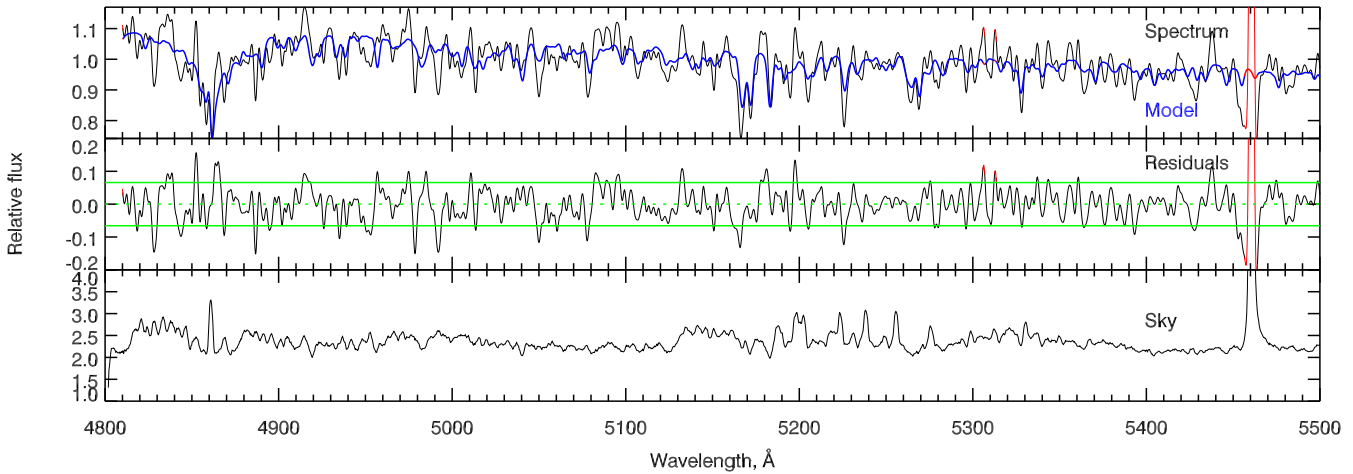}
\caption{ Two component fit of KDG\,64. In the upper panel with black we
plot the observations, smoothed with a Gaussian with a $\sigma\,=\,40$\,\kms{},
with blue we plot the best fitted model. 
In the middle panel we show the residuals (with black), smoothed in same way as
the observations, the one sigma error is in green. The masked pixels are
with red. In the last panel we show the sky. 
} 
\hskip 0cm
\label{fig:k64_fit}
\end{figure*}

\section{Discussion}

\subsection{The nature of the HII knot in KDG\,61}

According to our measurements presented above, there is a redshift difference 
of $344\pm6$\,\kms{} between the
stellar population of the galaxy and the \hii{} knot.
This knot (${\rm cz}\,=\,-123$\,\kms) is associated with the \hi{} cloud
detected at the position of the galaxy
\citep[][${\rm cz}\,=\,-140$\,\kms{}]{vanderhulst79,boyce01}.
Thus, because the velocity difference is one order of magnitude larger than 
the expected stellar velocity dispersion, we can conclude that the gas 
is not gravitationally bound to the main body of KDG 61.

KDG\,61 is spatially close to M\,81. We estimated its separation to be
about 40\,kpc. The velocity difference with M\,81 is 
$258\pm4$, adopting ${\rm cz} = -37\pm3$ \kms{} for M\,81 
\citep[HyperLEDA]{patu03}. Therefore, KDG\,61 is probably near the closest 
approach on an orbit highly inclined on the plane of the sky.
Such a pass could strip the gas from the galaxy, but it is very unlikely
that the \hi{} cloud and \hii{} knot were bound to KDG\,61 in
the past. It is well known that the central part of the M\,81 group is embedded
into the huge \hi{} cloud associated with the interaction between  M\,81, M\,82 
and NGC\,3077 galaxies. Thus, the most likely hypothesis is an
alignment between the gaseous and stellar structures on line of sight.

\citet{johnson97} discussed the nature of this emission region and stressed
its peculiarity. The ratio [OIII]5007/H$_\beta$ is as high as in AGN
and [NII]6584/H$_\alpha$. Thus, it appears to be in the transition region 
between \hii{} regions and AGNs on the \citet{baldwin81} diagnostic diagram.
\citet{johnson97} derived a metallicity of about 0.1 of the solar. It is confirmed 
recently by \citet{croxall09}. Assuming that this gas belongs to the galaxy,
the system would lie far out from the metallicity versus luminosity 
relation.
The oxygen abundance is too high for such a low mass object and it is close
to that of M\,81. This characteristic is shared by two other small objects of
the group: Garland and Holmberg\,IX.
Moreover, these objects lie in the regions of the \hi{} streams
connecting M\,81, M\,82 and NGC\,3077.
All these facts lead the authors to suggest that Garland, Holmberg~IX and
KDG\,61 are recently formed tidal dwarfs 
(Holmberg~IX and Garland were already considered as tidal dwarf galaxies 
by \citealt{vandriel98,makarova02,boyce01}). 
However, unlike other proposed tidal dwarfs (as Holmberg~IX, see \citealt{sabbi08})
KDG\,61 does contain a dominant old population.
In the Arp's loop, \citet{demello08} interpret the intermediate-age 
population as an ejection from the disk of M\,82 and M\,81 after the 
encounter. However, such an interpretation would not hold for KDG\,61, as the old
population  does not resemble a disk population. The tidal origin of KDG\,61
is confidently ruled out, and this galaxy appears to be a classical, bona-fide
dSph.

\citet{croxall09} also examines various possibilities for the nature
of this \hii{} knot. They rule out the cases of a supernova remnant
([Si{\sc ii}]6718/H$_\beta$ is too low) or planetary nebula (too extended).
The hypothesis of a wind blown superbubble is the most likely. The
detection of [He{\sc ii}]4686 in the spectrum may be the signature of
an ionization by Wolf Rayet stars. The horseshoe morphology of the region
at the ACS image, the presence of several blue stars ($\approx$10 Myr, OB association) and the
size of the region (major axis of 90 pc) strengthen this hypothesis.
The magnitudes integrated in a 4.5 arcsec aperture (80 pc) 
(corrected for Galactic extinction) are: $V_0 = 18.39\pm0.09$,
$I_0 = 18.96\pm0.07$ and $(V-I)_0 = -0.57\pm0.11$.  
The dereddened Galex magnitudes for the \hii{} knot are $m_{\rm NUV}^0 = 19.11$
and  $m_{\rm FUV}^0 = 19.15$ (we used the reddening coefficients from \citealt{seibert05}),
similar to the luminosity of the star forming complex observed 
in the \hi{} bridge between M\,81 and
M\,82 (Arp's loop) by \citet{demello08}.

\subsection{The nature of KDG\,61 and KDG\,64}

\begin{figure*}
\includegraphics[width=14cm]{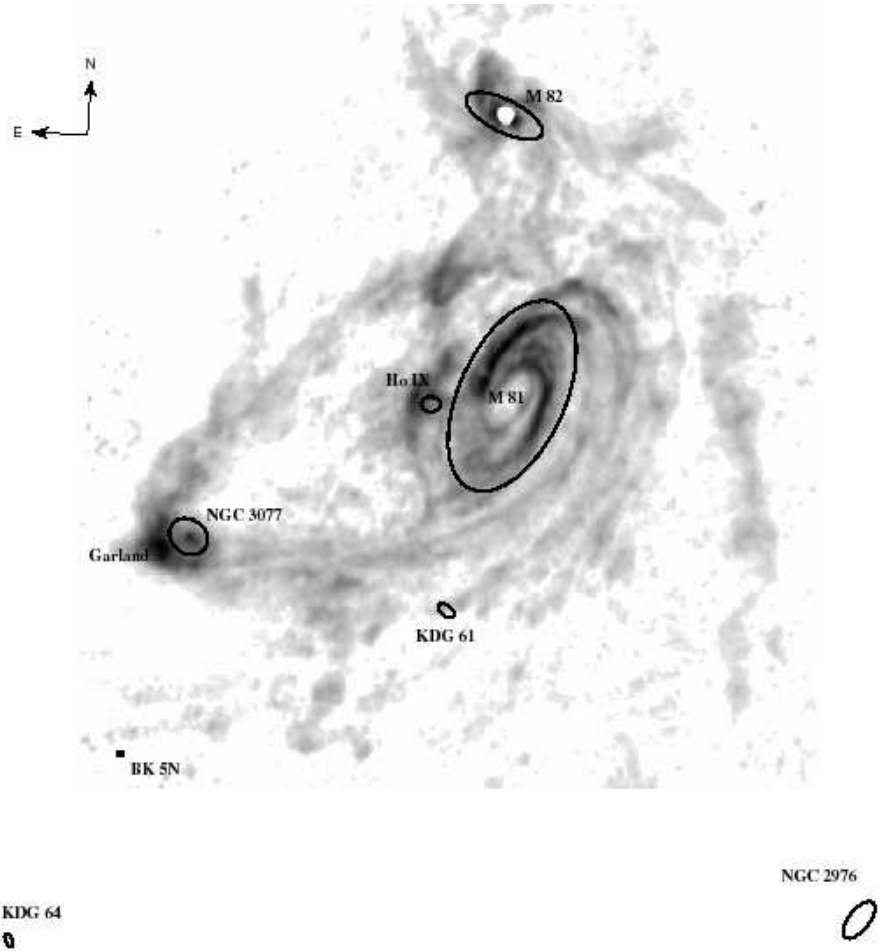}
\caption{Integrated HI map of the central part of the M~81 group 
from Yun et al. \citep{yun}. 
The ellipses indicate the galaxy sizes from \citet{CNG}.}
\label{fig:m81_gener}
\end{figure*}

The environment is often proposed to be responsible for the evolution
of the gas-rich small galaxies into dE/dSph through a transition-type
phase. The M\,81 group is an excellent site to catch these phenomena
in the act, due to the intense interactions between its members, as
seen from the filamentary \hi{} distribution (Fig. \ref{fig:m81_gener}).
and from the bridges connecting the principal galaxies
\citep{vanderhulst79,boyce01}.

Arguing on the presence of \hi{} gas, KDG\,61 was proposed in several occasions
to be such a transition type object \citep{johnson97,k00,boyce01}. But as we 
have seen above, this gas is not related to the stellar population.

\citet{boyce01} detected a spur of \hi{} extending south from NGC\,3077 to BK5N 
and KDG\,64. The optical position of KDG\,64 is close to a peak 
of \hi{} (see Fig. \ref{fig:m81_gener}) with the radial velocity of $-100$ \kms 
belonging to this feature.
The velocity of KDG\,64 is $-15$\,\kms.
The separation in velocity between the gas and the stars is not as large 
as in the case of KDG\,61, but a physical association between them is
likely not present, too. The velocity
difference between KDG\,64 and NGC\,3077 or M\,81 is small 
($\approx$ 30\,\kms{}) and the orbit lies probably close to the plan on the sky.
We can rule out KDG\,64 as being responsible for the tidal spur, which,
as suggested by \citet{boyce01} may be the result of the interaction
between BK5N and NGC\,3077.

Only the brightest part of stellar populations ($\sim$ 3 mag below TRGB)
were resolved at the HST/ACS images of KDG\,61 and KDG\,64. These are
numerous, predominantly old (about 12--14 Gyr) red giant branch stars and
intermediate age asymptotic giant branch stars, which indicate that
star formation episode occurred about 1--4 Gyr ago. 
At the distance of the M 81 group, the fainter stellar populations, 
like the horizontal branch and the lower part of the main sequence, 
are hardly resolved even with very deep exposures at the HST. 
Therefore, these populations are not
seen at the CMDs. Without these details it is difficult to resolve 
age-metallicity-SFR relation for the oldest ( $>6\div8$ Gyr) star formation
events, due to tight packing of the correspondent isochrones for the 
brightest part of the CMD. 
Integrated light spectroscopy brings here more reliable information.
In contrast to the M 81 group, dwarf galaxies in the Local Group can be resolved
into individual stars including the fainter populations.
According to the measurements made by \citet{dolphin05}, the Local Group dSphs 
in the same luminosity range as KDG\,61 and KDG\,64 ($M_B \sim -11 \div -13$ mag) 
reveal a dominant peak of star formation 10--13 Gyr ago 
with some residual star formation continuing until 1--3 Gyr.
It is consistent with our results.
A similar star formation history is also found in most of
the more massive dwarf galaxies $M_B\sim-16$ mag of the sample of \citet{kpr2009}, 
but, as variance with the present objects, these galaxies often display strong
radial gradients of age and metallicity.
It seems that the more luminous systems (2--3 mag brighter) have 
more resources for a second equally strong enhancing of star formation 
(about 3--6 Gyr ago). In less luminous systems, according to \citet{dolphin05}, 
we can expect one ancient burst of star formation of shorter duration and 
less metal enrichment. 

\section{Conclusions}

We have derived the star formation histories of two dE/dSph galaxies,
KDG\,61 and UGC\,5442 = KDG\,64, belonging 
to the nearby M\,81 galaxy group.

These galaxies have regular axisymmetric morphologies consistent
with an early-type classification. They appear to be dominated by
an old stellar population (approximately 12--14\,Gyr) of low metallicity
(${\rm [Fe/H]}\simeq-1.5$).
We also detected stars formed about 1 to 4\,Gyr ago in both galaxies
with a marginal metal enrichment.
KDG\,64 is slightly less luminous and it has a higher surface
brightness than KDG\,61, but both galaxies are nearly
a mid point of the luminosity sequence connecting the bright
dEs, like, for example NGC~205, and the smaller local dwarfs, like,
for example, Sculptor.

The well-known numerical simulations by \citet{yun99} gives
the time since nearest approach between M\,81--M\,82--NGC\,3077 of about 300 Myr ago. 
Recent star formation events in these and tidal dwarf galaxies
agree well with this age. But signs of earlier approaches and, therefore,
earlier common star formation events are probably ``erased'' by the recent 
interaction, and, therefore, we could not relate the recent star 
formation event in KDG\,61 and KDG\,64 (1--4 Gyr ago) to the large scale star
formation event in the M\,81 group.  

There is no sign of a gaseous component in none of the two galaxies
(\hi{} or ionized gas). On the basis of our radial velocity measurements 
on the stellar components, we reject previous suggestions for \hi{} clouds
association with the dwarfs.
We have also found, that the \hii{} knot previously suggested to belong to
KDG\,61 is in fact a wind-blown superbubble associated to a \hi{}
tidal stream projected on the line-of-sight.

This study was made with two complementary approaches: CMD fitting 
and full spectrum fitting. The first method uses CMDs 
usually obtained with HST/ACS,
while the last uses long slit spectra that can be acquired on a 
ground based telescope. The two methods are consistent and it is notable
that the full spectrum fitting can give finer constrains on the metallicity.
In full spectrum fitting, the metallicity is directly constrained by the metal
absorption lines from the stars, while CMDs rely on broad bands colours
and on stellar evolution models. It is the first time that a detailed
spectroscopic determination of the SFH was successfully applied to
low surface brightness galaxy (with low S/N data).

We like to notice the agreement between the two approaches.
It is promising for the future where both methods will be applicable within
a distance of 20\,Mpc. The spectroscopic analysis of low surface brightness
objects may also be used, at moderate cost, in a large volume of
the local universe to explore the diversity of SFHs, which can be related to
other characteristics of the galaxies and to the environment.

\section*{Acknowledgements}
We are thankful to Olga Sil'chenko and Victor Afanasiev for 
help with observation. DM, LM and MK thank Observatoire de Lyon 
for the invitation and hospitality. The work was partially 
supported by RFBR grants 08--02--00627 and 07--02--00792.
MK was supported by DO 02-85/2008 of Bulgarian Scientific 
Research Fund and Programa Nacional de Astronom\'{\i}a y
Astrof\'{\i}sica of the Spanish Ministry of Science and 
Innovation under grant \emph{AYA2007-67752-C03-01}. 

\bibliographystyle{mn2e}
\bibliography{dsph}   

\appendix

\section[]{Background galaxy projected on KDG\,64}
\label{sect:bg}

\begin{figure*}
  \includegraphics[height=0.33\textheight]{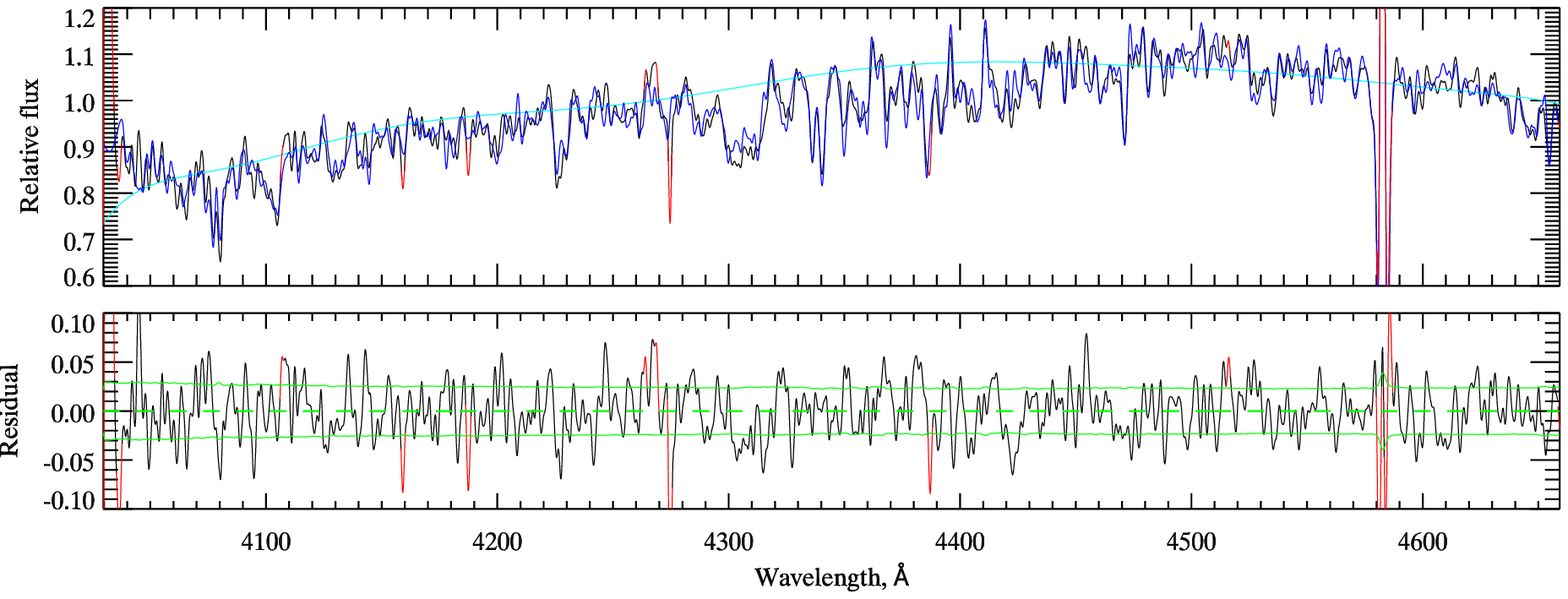}
  \caption{Five component fit (see the text) of the background galaxy.
    blue is the best fitted model, cyan is the multiplicative polynomial,
    black the objected smoothed with Gaussian sigma of 40\,\kms. 
    In the lower panel we plot the residuals from the fit (in black, a
    gain smoothed in same way as
    the observations) and the one sigma error in green. The masked pixels are
    with red. 
  } 
\hskip 0cm
\label{fig:bg_fit}
\end{figure*}

To correct the residual contamination in KDG\,64 from the background 
galaxy, we adopted the following strategy:
\begin{itemize}
\item{We prepared two extractions, one containing mostly KDG\,64 (the
  regions marked with blue horizontal lines on Figure\,\ref{fig:k64prof}) and 
  one containing mostly the background object (the grey part on 
  Figure\,\ref{fig:k64prof}). We named them extraction1 and extraction2,
  respectively.}
\item{We decomposed the profile along the slit to a Gaussian (for the background
  galaxy), plus a polynomial (for KDG\,64). From this decomposition, we measured
  that extraction2 should be contaminated by KDG\,64 by 80\,per\,cents. The 
  contamination of the background galaxy in extraction1 should be less than 
  1\,per\,cent.}
\item { At first we attempted to fit extraction2 with a two component
  model:  one SSP for KDG\,64 and one SSP for the background galaxy.
  However, we realized that we cannot fit the background galaxy
  (extraction2) with a SSP. We adopted a five component model
  (Fig.\,\ref{fig:bg_fit}): 1 SSP for KDG\,64, 2 SSPs for the
  background galaxy and 2 delta functions for  ($H_{\gamma}$ and
  $H_{\delta}$). We set the limits for the background SSPs  to be
  between 0.001 and 0.5\,Gyr and 5 and 16\,Gyr. The emission lines
  were broadened with the LOSVD of the galaxy. We found an young
  population of 0.016\,Gyr with -1\,dex of metallicity, weighting
  15\,per\,cents in light and 1\,per\,cent in mass  and an old
  population of about 7\,Gyr, super-solar metallicity,  23\,per\,cents
  in light and 94\,per\,cents in mass (the rest of the  light is going
  to KDG\,64 component).  Our determination of radial velocity of the
  background galaxy  (${\rm cz}=57540$\,\kms), disagrees with the
  value published by  \citet[][${\rm cz}=46530$\,\kms]{sharina01}.
  After checking the plotted spectrum in \citet{sharina01},  we
  conclude that the published value is an error (a typo or wrong
  conversion). Our estimation of velocity dispersion 
  for the background galaxy is  $\sim130$\,\kms.}
\item{As a last step we used the best fit of the background galaxy to analyse
  extraction1. When using single burst decomposition for KDG\,64 we found
  less than 1\,per\,cent of the light going to the background object, 
  consistent with the profile decomposition.}
\end{itemize}

\label{lastpage}

\end{document}